\documentclass[11pt]{article} 
\usepackage[left=0.75in,top=1in,right=2in,bottom=1in]{geometry} 

\usepackage{graphicx}
\usepackage[table]{xcolor}
\usepackage{textgreek}
\usepackage[font={small,sf}]{caption}
\usepackage{amssymb,amsfonts,amsmath,amssymb}
\usepackage[numbers,sort&compress]{natbib}
\RequirePackage{hyperref}

\usepackage{libertine}
\usepackage[libertine]{newtxmath}

\begin{document}

\title{\bf Dislocation evolution during plastic deformation: \\ Equations vs. discrete dislocation dynamics study} 
\author{Kamyar M. Davoudi$^{a,b}$\footnote{corresponding author: {\tt davoudi@seas.harvard.edu} and {\tt davoudi@alumni.harvard.edu}} \  and   Joost J. Vlassak$^a$}
\date{\footnotesize $^a$ School of Engineering and Applied Sciences, Harvard University, Cambridge, MA 02138, USA \\
$^b$ Department of Chemical and Materials Engineering, University of Alberta, Edmonton, AB, Canada T6G 1H9}

\maketitle 
 
\begin{abstract}

{\sf Equations for dislocation evolution bridge the gap between dislocation properties and continuum descriptions  of plastic behavior of crystalline materials. Computer simulations can help us verify these evolution equations and find their fitting parameters. In this paper, we employ discrete dislocation dynamics to establish a continuum-based model for the evolution of the dislocation structure in polycrystalline thin films. Expressions are developed for the density of activated dislocation sources, as well as dislocation nucleation and annihilation rates. We demonstrate how size effect naturally enters the evolution equation. Good agreement between the simulation and the model results is obtained. The current approach is based on a two-dimensional discrete dislocation dynamics model, but can be extended to three-dimensional models.

\vspace{\baselineskip}
\noindent{\bf keywords:} Dislocation Evolution, Nucleation Rate, Annihilation Rate, Density of  Activated Sources,  Size Effect, Discrete Dislocation Dynamics }
\vspace{2\baselineskip}
\end{abstract}


\section{Introduction}

``It is sometimes said that the turbulent flow of fluids is the most difficult remaining problem in classical physics. Not so. Work hardening is worse'', remarked Cottrell \cite{Cottrell2002}. Work hardening, a mechanism that occurs in crystalline metals, manifests as a rise in the stress required for continued plastic deformation. Despite all the efforts that have been put into the study of work hardening in the past 80 years, there is currently no generally accepted theory explaining all aspects of it \cite{Kubin2009}; finding a theory of work hardening is now as hopeless as ever, and research is aimed at establishing a model instead \cite{Kocks2003}. 

Plastic deformation is path dependent. Plastic deformation is an irreversible, highly dissipative process that occurs far from equilibrium \cite{Kubin1992,Sauzay2011}; thermodynamic extremum principles are not applicable \cite{Hillert2006,Sauzay2011}.  In addition, the characteristic time and length scales that are involved in the physics of the processes in the dislocation core and in the deformation of bulk materials differ by several orders of magnitude.  Evolution equations can serve as a bridge between elementary dislocation properties and a continuum model for plastic deformation.  Although many equations of dislocation evolution have been developed  for materials in bulk form \cite{Kocks1975,Mecking1981,Estrin1984, Walgraef1985b,Walgraef1985,Walgraef1985a,Aifantis1986,Follansbee1988,Hahner1996a,Hahner1996,Nix2011a}, there have been few attempts \cite{Devincre2008,Ryu2013a} to use computer simulations  to check the validity of these equations or find the fitting parameters of those equations. Furthermore, the evolution equations developed for materials in bulk need to be adjusted for thin films and other small structures.

Of the various computer simulation techniques, discrete dislocation dynamics (DDD) is the most suitable method to model thin films and many small structures at the micron scale and below. In this method, the material is modeled as a continuum that contains dislocations. Grain boundaries may also be included for polycrystalline materials. Dislocations nucleate, move and are destroyed under a few simple constitutive laws. Three-dimensional DDD models capture the physics of problems more accurately than 2D models, but they are computationally demanding and are not easily applied to thin films. Therefore, most three-dimensional models are limited to single crystals, very small strains, small volumes of material, and low dislocation densities. For example, ParaDis, a powerful three-dimensional DDD code, which was originally developed at the Lawrence Livermore National Laboratory, can only model single-crystal materials under simple boundary conditions.  A very recent time integration algorithm proposed by Sills et al. \cite{Sills2016} can speed up 3D calculations, and may make achieving more realistic strains possible. Two-dimensional discrete dislocation dynamics models, on the other hand, can model complicated boundary value problems, polycrystalline materials, realistic dislocation densities, and relatively large strains with much less computational effort.  Some additional features of 3D models such as junction formation and destruction can be incorporated into 2D approaches through an extra set of constitutive rules in what is often called 2.5 D models \cite{Benzerga2004a, Gomez-Garcia2006, Keralavarma2016}. While 2D models necessarily miss some of the physics, recent studies have shown that two- and three-dimensional simulations predict remarkably similar results in some cases \cite{Ispanovity2010,Ispanovity2013}.  For these reasons, both types of simulations are popular and are employed to study different phenomena (see, e.g., \cite{Agnihotri2015,Keralavarma2015,Shishvan2016a,
Gurrutxaga-Lerma2017,Quek2016,El-awady2015a,Madec2017}).

In this paper, we derive a continuum model for the dislocation evolution in polycrystalline thin films that are passivated on both surfaces and use discrete dislocation simulations to verify the model. While in most DDD analyses edge dislocations can only glide, dislocations in this study are allowed to both glide and climb. Dislocation climb is a mechanism by which edge dislocations trapped at glide barriers can leave their primary slip planes. Thus climb acts as a softening process and may be taken as representative of a range of softening mechanisms that occur in a material.

The paper is organized as follows; first the framework of the two-dimensional DDD model is briefly described. Then we derive an expression for the density of activated dislocation sources. The next sections are devoted to deriving expressions for dislocation nucleation and annihilation. Combining these relations, a governing equation for the dislocation evolution is proposed in the final section.

\section{Discrete Dislocation Model}

In discrete dislocation dynamics, a material is modeled as an elastic solid that contains dislocations. Simulations are then carried out in an incremental fashion. At a given instant of time, it is assumed that the material is in equilibrium and that the displacement and stress fields are known. An increment of strain is prescribed and the positions of the dislocations in the material, the displacement field, and the stress field are updated using the following procedure: (1) The Peach-Koehler force on each dislocation is calculated; (2) in response to the Peach-Koehler forces, the dislocation structure evolves: dislocations move, new dislocations nucleate, and others are annihilated; (3) the stress state in the solid is calculated for the updated dislocation arrangement. Steps 1 and 3 follow from elasticity; step 2 requires the formulation of constitutive rules for dislocation behavior. To determine the stress state at each time step,  we follow the superposition procedure proposed by Van der Giessen and Needleman \cite{VanderGiessen1995}. According to this procedure, the elastic fields are written as the superposition of two fields: one field due to the dislocations in an infinite medium and an image field that enforces the boundary conditions.  When the local shear stress on a dislocation source inside the material exceeds the nucleation strength of the source for a specific time, the source emits a dislocation dipole. The distance between the two dislocations is taken such that the attraction between the two dislocations in the dipole is balanced by the source strength. When two dislocations of opposite sign on the same glide plane opposed each other within a critical distance, say 6$b$ where $b$ is the magnitude of the Burgers vector, the dislocations annihilate each other and are removed from the model. At temperatures above 20 K, phonon drag is large enough to make dislocations quickly reach the overdamped regime \cite{Kubin1992a,Davoudi2017,Bulatov2006a} and a linear relationship between the Peach-Koehler force on a dislocation and its glide velocity is assumed. 

If a non-vanishing normal stress exerts a force on a dislocation perpendicular to its glide plane, the dislocation starts to climb by emission or absorption of vacancies, resulting in a local change in the vacancy concentration. Under most experimental conditions \cite{Messerschmidt2010}, a quasi steady state develops, in which the dislocation climb velocity is controlled by the flux of vacancies. The vacancy flux, in turn, is determined by the gradient of the vacancy concentration (or more accurately by the gradient of the chemical potential) \cite{Weertman1955,Schoeck1957,Schoeck1980}. Assuming that the concentration of vacancies outside a cylinder of radius $R$ around a straight dislocation is equal to the equilibrium concentration without stress, the climb velocity reads \cite{Schoeck1957,Schoeck1980}
\begin{equation}
v_c=\frac{2\pi D_0}{b \ln(R/b)}\exp\left(-\frac{\Delta E_{sd}}{k_B T}\right)\left[\exp\left(\frac{F_c b^2}{k_B T}\right)-1\right],
\end{equation}
where $\Delta E_{sd}$ is the vacancy self-diffusion energy, and $D_0$ the pre-exponential diffusion constant. Because $R/b$ appears as the argument of a logarithm, its precise value has little influence on the climb velocity, and is often taken equal to $2\pi$. The mechanical climb force per unit length, $F_c$, is taken positive when it favors vacancy emission. Because the climb velocity is typically much smaller than the glide velocity, different time steps are used for climb and glide. In this paper, the time step for climb is taken 100 times larger than the time step for glide.

To find the correct displacement field due to a dislocation dipole where one of the dislocations climbs from $(x_0,y_0)$ to $(x_0,y_1)$, the following terms need to be added to the $x$-component of the displacement field published in most texts on dislocations:
\begin{align*}
\frac{b}{2\pi }\Big[ {{\tan }^{-1}}\left( \frac{y-{{y}_{1}}}{x-{{x}_{0}}} \right)-{{\tan }^{-1}}\left( \frac{y-{{y}_{0}}}{x-{{x}_{0}}} \right)
+{{\tan }^{-1}}\left( \frac{x-{{x}_{0}}}{y-{{y}_{1}}} \right)-{{\tan }^{-1}}\left( \frac{x-{{x}_{0}}}{y-{{y}_{0}}} \right) \Big].
\end{align*}
These extra terms account for the displacement caused by the emission or absorption of vacancies during climb \cite{Davoudi2012,Davoudi2014,Ayas2012a,Ayas2015}, and  ensures that the displacement discontinuities associated with dislocation motion occur along the path of the dislocations.

\begin{figure}
\centering
\includegraphics[width=8cm]{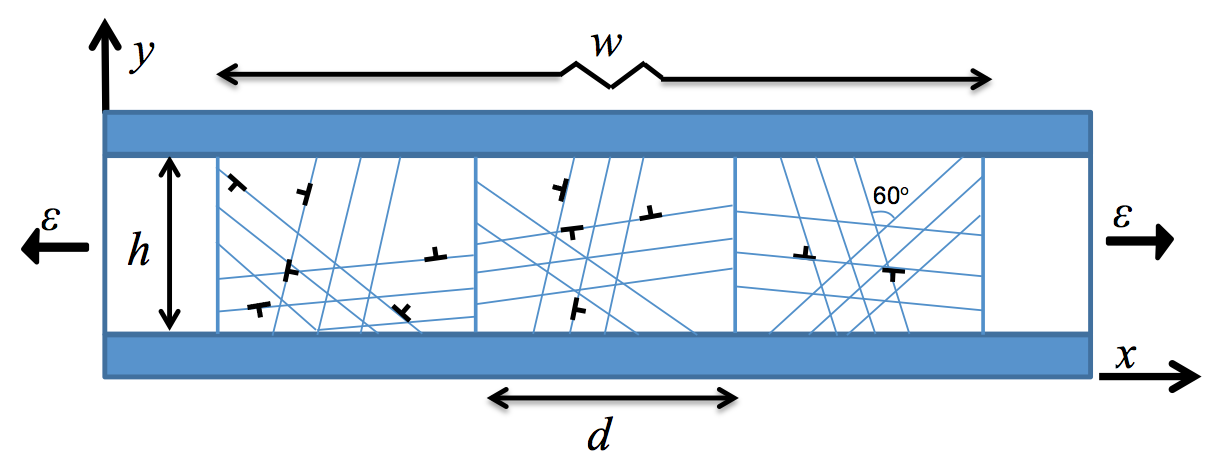}
\caption{{\label{Fig:Geometry}}Schematic representation of the thin-film model}
\end{figure}

Discrete dislocation dynamics simulations were performed for freestanding polycrystalline aluminum films passivated on both surfaces. The films were subjected to uniaxial tension as illustrated schematically in Fig.~\ref{Fig:Geometry}. Thin films of aluminum often have a columnar grain structure, which was modeled as a two-dimensional array of randomly oriented rectangular grains of thickness $h$, in line with Nicola et al. \cite{Nicola2006}. The calculations were carried out for a unit cell of width $w$ consisting of six grains of uniform size $d$ ($w=6d$). Each grain had three sets of slip planes that differed by 60$^\circ$ \cite{Rice1987}.  The grain size of the film was 1 \textmu m, while the thickness of the passivation layers was taken to be 20 nm. The passivation layers were assumed to deform elastically and had the same elastic properties as the film material. Both grain boundaries and passivation layers were assumed impenetrable to dislocations. Periodic boundary conditions were applied at the left and right boundaries of the model. Plane-strain conditions were assumed in the $xy$-plane; the tensile stress in the film was calculated as the stress $\sigma_{xx}$ averaged over the thickness of the film.

The films were initially dislocation free, but dislocation sources were randomly distributed on the slip planes. The density of dislocation sources was taken as 15 \textmu m$^{-2}$ in all simulations. The nucleation strengths of the dislocation sources, $\tau_{\rm nuc}$, were randomly selected from a normal distribution with a mean of  $\tau_{\rm avg}=100$ MPa and a standard deviation of $\tau_{\rm sd}=20$  MPa. To limit computation time, a high strain rate of 4000 s$^{-1}$ was used in all simulations.  All simulations were performed at a temperature of 900 K. Using molecular dynamics, two- and three-dimensional DDD simulations, and  existing experimental results, Davoudi \cite{Davoudi2017}  recently studied how DDD parameters change with the temperature, and how these variations affect the yield strength of aluminum thin films. The choice of other parameters and the model are outlined in more detail in Ref. \cite{Davoudi2012,Davoudi2014}. 

To reduce the effects of the initial conditions, four realizations of the model were run for each set of parameters. Realizations differed from each other with respect to grain orientations and locations of dislocation sources.

\section{Nucleation and Annihilation Rates}

G.I. Taylor \cite{Taylor1934} linked the shear strength of a material to its microstructure. He recognized that the (athermal) flow stress is proportional to the square root of the dislocation density (see Appendix \ref{Section:TaylorEquation}).  To be useful, however, this equation requires knowledge of how the dislocation density and structure evolve during plastic deformation of the material. In the absence of a free surface, the evolution of the dislocation density depends on two simultaneous mechanisms: dislocation nucleation and annihilation. This section is devoted to the derivation of relations that describe the generation and annihilation of dislocations. Simple expressions are developed and compared with simulation results. 
 
\subsection{Density of Activated Sources}{\label{Section:ActiveSources}}

The dislocation nucleation rate is proportional to the density of activated  dislocation sources in a material: the higher the density of activated  sources, the higher is the generation rate. Here we evaluate the density of  activated  sources as the stress in the film increases.

Let the probability distribution function (PDF) of the source strengths be denoted by $\phi(\tau)$. Then, if the shear stress on a dislocation source is $\tau'$, the probability that $\tau'$ is larger than the source strength and makes the source activated is given by  $\Phi(\tau')=\int_{\infty}^{\tau'} \phi(\tau) {\rm d}\tau$. In this paper, $\phi(\tau)$ is a normal distribution with a mean $\tau_{\rm avg}=100$ MPa and a standard deviation $\tau_{\rm sd}=20$ MPa; thus, its cumulative distribution function (CDF) reads 
\[\Phi(\tau)=\frac{1}{2}\left[1+{\rm erf}\left(\frac{\tau-\tau_{\rm avg}}{\sqrt{2} \tau_{\rm sd}}\right)\right],\]
where ``erf'' is the error function. Now suppose $\tau_i$ is the resolved shear stress on a slip system. If we ignore local shear stress variations, and assume the number of sources on each slip system is approximately the same, the density of the sources that have been activated $\rho_{\rm source}$ can be estimated as 
\begin{align}\label{Eqn:PlanesSigma}
\begin{split}
\rho_{\rm source}&=c \sum_{i=1}^{N_{\rm slip\,sys}}\frac{\rho_{\rm source}^0}{N_{\rm slip\, sys}}\Phi(\tau_i)\\
&  = c \frac{\rho_{\rm source}^0}{2}\left[1+\frac{1}{N_{\rm slip\,sys}}\sum_{i=1}^{N_{\rm slip\,sys}} {\rm erf}\left(\frac{\sigma \sin \theta_i |\cos \theta_i|-\tau_{\rm avg}}{\sqrt{2}\tau_{\rm sd}}\right)\right]           
\end{split}
\end{align}
where $N_{\rm slip\,sys}$ is the number of slip systems, $\rho_{\rm source}^0$ is the density of all dislocation sources (activated  or not) in the film,   $\theta_i$ is the angle between  slip system $i$ and the $x$-axis, and $c$  is a proportionality constant of order unity.

In Eq.~\eqref{Eqn:PlanesSigma}, $\rho_{\rm source}^0$ and $N_{\rm slip\ system}$ are fixed, and the only variable is the resolved shear stress $\tau_i$.   As the number of realizations increases, the average  of $\rho_{\rm source}$ becomes independent of the choice of the slip orientations and approaches the following integral:
\begin{align}\label{Eqn:IntegralPlanesSigma}
\begin{split}
c \frac{\rho_{\rm source}^0}{2}\Bigg[1+\frac{1}{\pi} \int_0^{\pi} {\rm erf}\left(\frac{\sigma \sin \theta \left|\cos \theta\right|-\tau_{\rm avg}}{\sqrt{2}\tau_{\rm sd}}\right)  {\rm d}\theta \Bigg].
\end{split}
\end{align}

\begin{figure}[h!]
\centering
\begin{tabular}{p{0.1cm} p{8cm}}
\vspace{10pt}{(a)} & \vspace{0pt}\includegraphics[width=8cm]{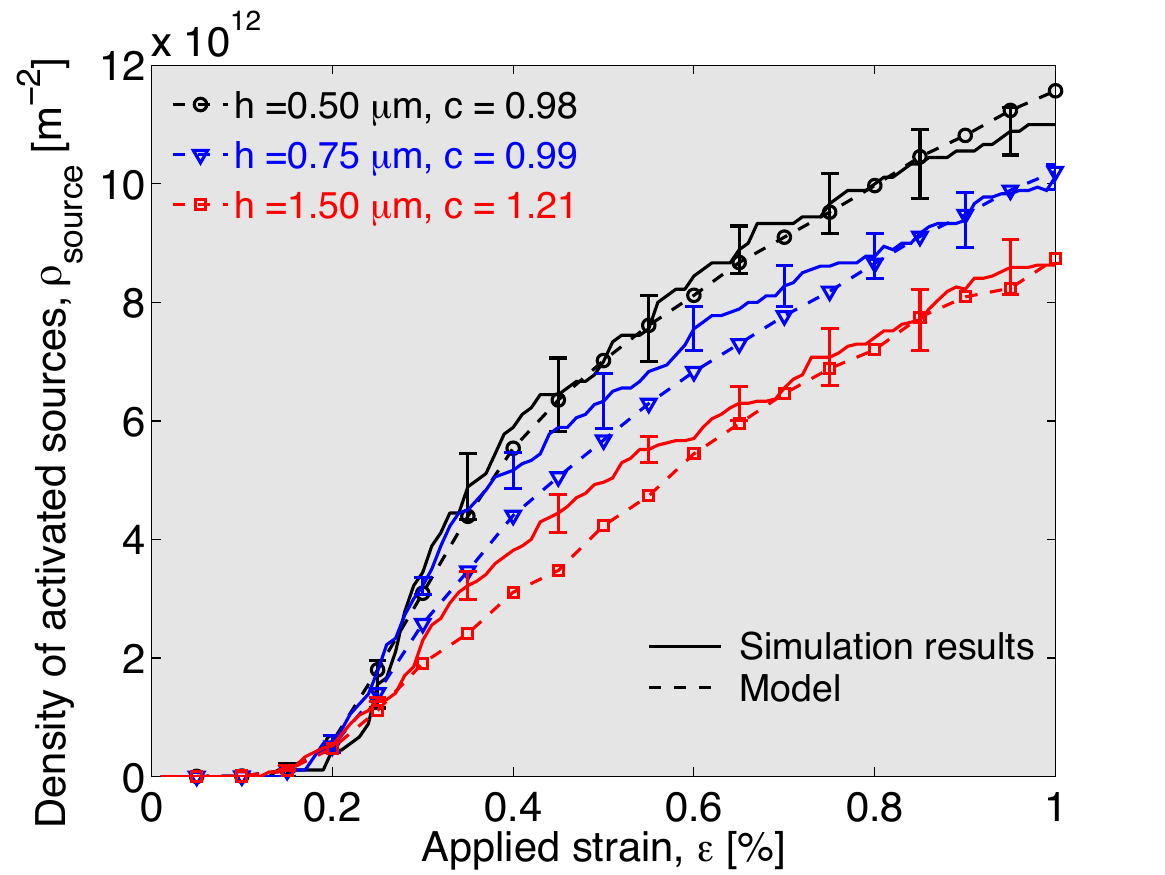} \\
\vspace{10pt}{(b)}&  \vspace{0pt}\includegraphics[width=8cm]{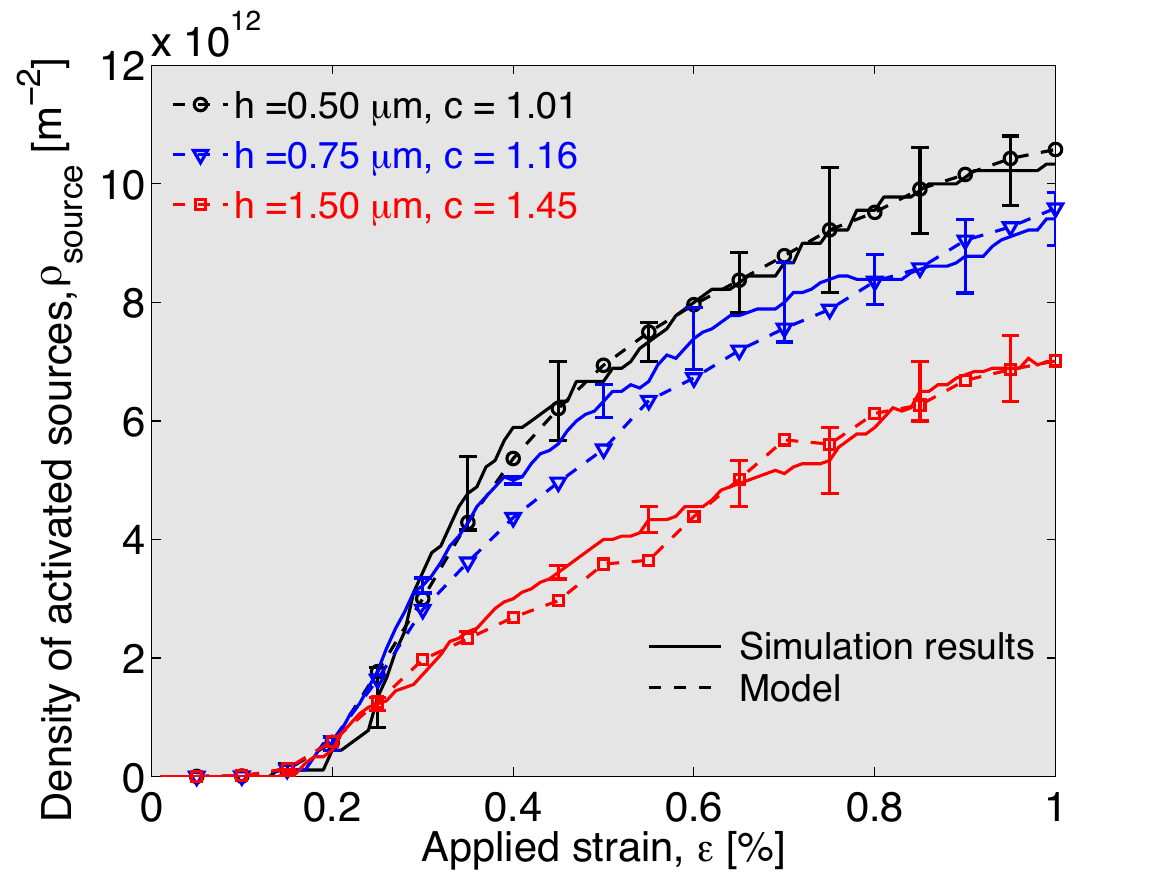} 
\end{tabular}
 \caption{\label{fig:ActiveSources}The solid curves denote the average density of activated  dislocation sources in the simulations for three different film thicknesses $h$ when dislocations (a) can only glide or (b) can glide and climb. The dashed curves are the results of the procedure described in Section~\ref{Section:ActiveSources}. Error bars show the standard error.}
\end{figure}

Figure~\ref{fig:ActiveSources} shows as a function of applied strain the density of activated  sources for three different DDD simulations. The same figure also shows the density of activated  dislocation sources determined from Eq.~\eqref{Eqn:PlanesSigma}, where the resolved shear stress was determined from the average normal stress obtained in the simulations and where the proportionality constant was treated as a fitting parameter.  The figure illustrates that Eq.~\eqref{Eqn:PlanesSigma} provides a good description of the activated  dislocation density and that the fitting parameters are all close to unity. The error in this approximation arises from three different sources: (1) Use of the strength distribution instead of the actual strength of a source introduces an error that decreases with increasing sample size and increased number of dislocation sources in the model. (2) The assumption that the number of sources is the same on each slip system also causes an error that decreases with the number of sources in the DDD model and thus better statistical sampling. (3) The main error  arises from using the average normal stress to calculate the resolved shear stress instead of the local stress, which depends on the local dislocation configuration. These errors are captured by the proportionality constant $c$ and cause the constant to deviate from unity. 

\subsection{Dislocation Nucleation}{\label{Section:Nucleation}}

An activated  dislocation source will emit a new dislocation whenever the previously emitted dislocations have traveled far enough to decrease the back stress on the dislocation source. If that travel distance is denoted by $y_{\rm back}$, then the rate at which dislocations are generated $\dot{\rho}_+$ is given by \cite{Kocks1975}
\begin{equation}{\label{Eq:rhopos1}}
\frac{{\rm d}\rho_+}{{\rm d} t}=\rho_{\rm source}\frac{\bar v}{y_{\rm back}}.
\end{equation}
Here $\rho_{\rm source}$ is the density of activated  sources, and $\bar{v}$ is the average dislocation velocity, which is related to the plastic shear strain rate by Orowan's equation $\dot{\gamma}_p=\rho b \bar{v}$. 

The back stress on the source due to a dislocation that has traveled a distance $y$, is proportional to $\mu b/(2\pi y)$. If the difference between the resolved applied  stress, $\tau$, and this back stress exceeds the nucleation stress $\tau^*$, then a dislocation dipole is generated. Note that $\tau^*$ is the nucleation strength of the first source emitting dislocations at the onset of plastic deformation.  The minimum distance required for continuation of dislocation emission, $y_{\rm back}$,  can be obtained from the following stress balance:
\begin{equation}{\label{Eq:yback}}
\tau-k \frac{\mu b}{2\pi y_{\rm back}}=\tau^*,
\end{equation}
where $k$ is a constant. The resolved shear stress, $\tau$, and the applied normal stress for a polycrystal, $\sigma$, are related through $M\tau=\sigma$, where $1/M$ is the average Schmid factor \cite{Kubin2013}. Also the relationship between shear and normal plastic strains is given by $\gamma_p=M\varepsilon_p$. In this model, $M=3.1$, which is the same as the Taylor factor for fcc and bcc materials.  If we substitute $\bar{v}$ and $y_{\rm back}$ from Orowan's equation and Eq.~\eqref{Eq:yback}, respectively, into Eq.~\eqref{Eq:rhopos1}, we can then use the relationships between shear and normal stresses and strains to arrive at the following equation:
\begin{equation}{\label{Eq:rhopos}}
\frac{{\rm d}\rho_+}{{\rm d} t}=2\pi\beta_1\rho_{\rm source}\frac{\sigma-\sigma^*}{\rho \mu b^2}\frac{{\rm d} \varepsilon_p}{{\rm d} t},
\end{equation}
where $\beta_1$ is a proportionality constant, and $\sigma^*$ is the applied stress at the onset of plastic strain. 

\begin{figure}[h!]
\centering
\begin{tabular}{p{0.1cm} p{8cm}}
\vspace{10pt}{(a)}&\vspace{0pt} \includegraphics[width=8cm]{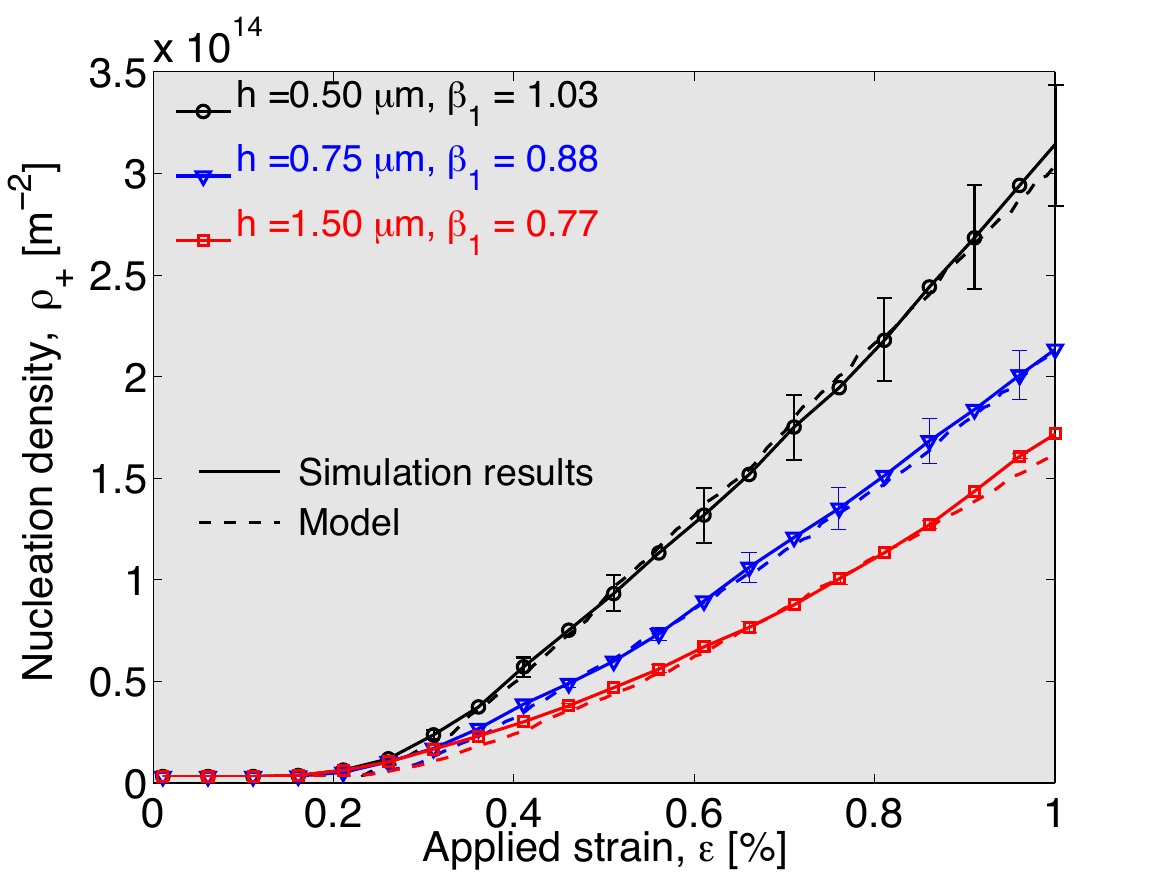} \\
\vspace{10pt}{(b)}&\vspace{0pt} \includegraphics[width=8cm]{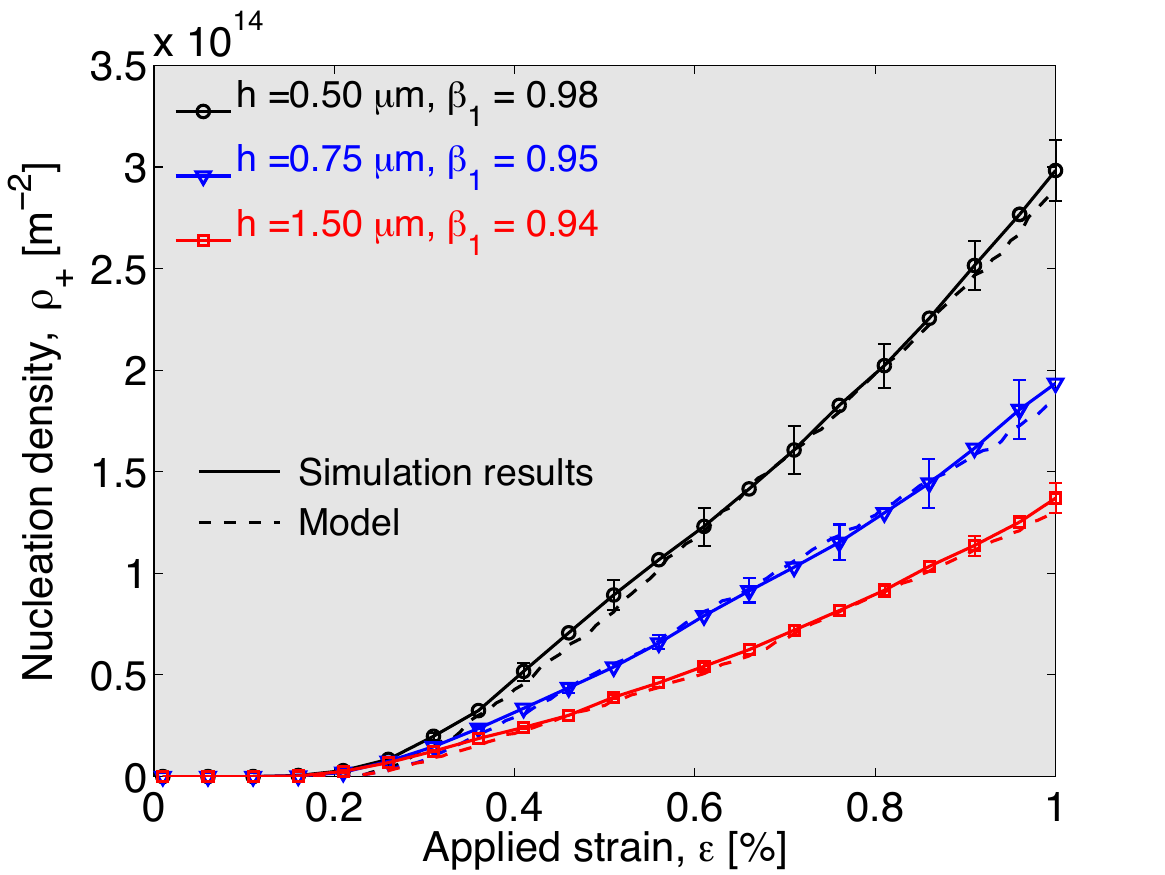} 
\end{tabular}
\caption{\label{Fig:rhopos} This figure shows how the dislocation density associated with nucleation, $\rho_+$, changes with applied strain, (a) for glide only and (b) for glide and climb. The solid lines were obtained from the simulations for three different film thicknesses $h$. The dashed lines were obtained from the model developed in this paper. Error bars show the standard error.}
\end{figure}

Figure~\ref{Fig:rhopos}  shows the density of the dislocation nucleation (cumulative number of nucleated dislocations per unit area) versus applied strain obtained from the simulations, denoted by solid lines. The nucleation density can also be determined by integrating Eq. ~\eqref{Eq:rhopos} using the stress, plastic strain, and dislocation density from the simulations and by considering $\beta_1$ as a fitting parameter.  The results are shown as the dashed curves in Fig.~\ref{Fig:rhopos}. Evidently Eq.~ \eqref{Eq:rhopos} provides a very good description of the nucleation rate.

\subsection{Dislocation Annihilation}{\label{Section:Annihilation}}
The dislocation annihilation rate is inversely proportional to the mean free path before a dislocation encounters a dislocation of opposite sign. As the mean free path $y_m$ decreases, the annihilation rate increases. Furthermore, the higher the dislocation density, the larger the chance of annihilation. Thus, the annihilation rate can be written as 
\begin{equation}{\label{Eq:annihilation}}
\frac{{\rm d}\rho_-}{{\rm d} t}\propto\rho \frac{\bar{v}}{y_m},
\end{equation}
as suggested by Kocks et al. \cite{Kocks1975}.  From geometry, the dislocation density and the mean free path are related through $\rho^{-1}=z_m\,y_m$, where $z_m$ is the average distance between dislocations in the direction perpendicular to the slip planes. Assuming that the activated  dislocation sources are randomly distributed, $z_m^{-1}$ is proportional to $2\sqrt{\rho_{\rm source}}$. Substituting these expressions into Eq. \eqref{Eq:annihilation} and using Orowan's equation, the annihilation rate becomes
\begin{equation}{\label{Eq:rhoneg1}}
\frac{{\rm d}\rho_-}{{\rm d}t}=\beta_2 \frac{M\rho}{2b\sqrt{\rho_{\rm source}}}\frac{{\rm d}\varepsilon_p}{{\rm d} t},
\end{equation}
where  the relationship $\gamma_p=M\varepsilon_p$ has been used to convert shear strain rates into normal strain rates and where $\beta_2$ is a dimensionless constant. 

At the onset of plastic deformation when the dislocation density is low, the assumption that $y_m$ is proportional to $2\sqrt{\rho_{\rm source}}/\rho$  may yield a value that is larger than the length of the slip plane in very thin films. In this case, the mean free path is solely determined by geometry and may be taken proportional to the film thickness $h$. The annihilation rate then becomes
\begin{equation}{\label{Eq:rhoneg2}}
\frac{{\rm d}\rho_-}{{\rm d}t}= \frac{\beta'_2 M}{bh}\frac{{\rm d}\varepsilon_p}{{\rm d} t},
\end{equation}
where $\beta_2'$ is another dimensionless constant. There may be different dislocation annihilation regimes during plastic deformation of very thin films. At the onset of plastic flow when the dislocation density in the film is low, Eq. \eqref{Eq:rhoneg2} may be valid, but as more and more dislocations are generated,  the mean free path decreases and Eq.~\eqref{Eq:rhoneg1} applies. 

\begin{figure}[h!]
\centering
\begin{tabular}{p{0.1cm} p{8cm}}
\vspace{10pt}{(a)} 
&\vspace{0pt}\includegraphics[width=8cm]{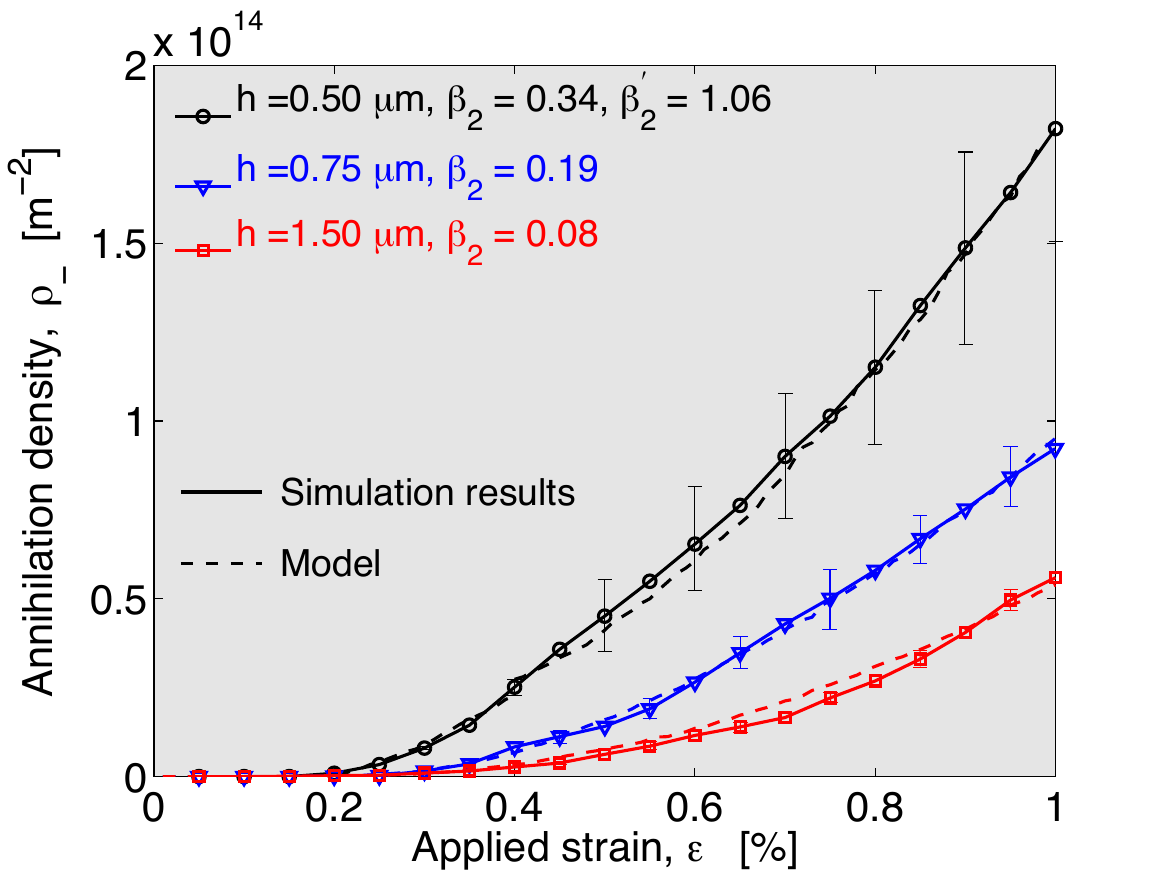} \\
\vspace{10pt}{(b)}
&\vspace{0pt}\includegraphics[width=8cm]{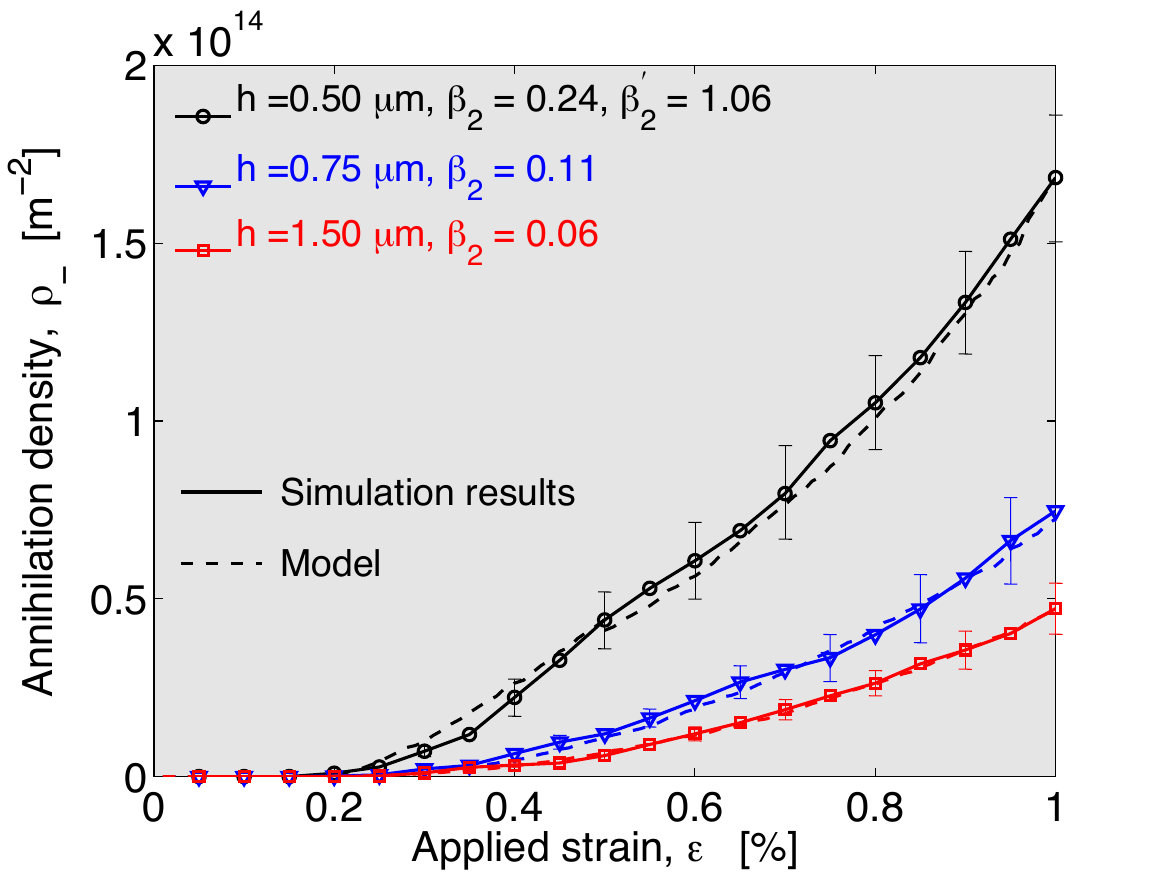} 
\end{tabular}
\caption{\label{Figure:rhoneg_final} This figure shows how density of dislocation annihilation changes with the applied strain (a) for glide only and (b) for glide and climb. The solid lines were obtained from the simulations for three different film thicknesses $h$. The dashed lines are the results of the model presented in this paper. Error bars show the standard error.}
\end{figure}

Figure~ \ref{Figure:rhoneg_final} shows how the dislocation annihilation density (cumulative number of annihilated dislocations per unit area) varies during plastic deformation of a thin film. The solid curves represent the annihilation density obtained from discrete dislocation simulations, while the dashed curves represent the results obtained from the model -- Eq.~\eqref{Eq:rhoneg1} in most cases; Eq.~\eqref{Eq:rhoneg2} had to be used only for initial flow of the thinnest film. The model provides a very good fit to the simulation results in all cases. The values of the $\beta_2$ coefficients are quite small and decrease with increasing film thickness. This happens because only dislocations of opposite signs annihilate each other, and the distance between positive and negative dislocations becomes larger with increasing film thickness, thus reducing the probability of annihilation. The values of the coefficients also decrease when dislocation climb is enabled, primarily because climb tends to disperse dislocations, thus decreasing the probability of annihilation. The value of $\beta'_2$ , on the other hand, seems independent of whether dislocations climb, because dislocation climb only becomes significant at high stresses where the mean free path is smaller than the length of the slip planes in the films.

\subsection{Evolution of the dislocation density}\label{Section:EvolutionEqution}

The evolution of the dislocation density depends on both the dislocation nucleation and anihilation rates: $\dot{\rho}=\dot{\rho}_+-\dot{\rho}_-$. Combining Eqs.~\eqref{Eq:rhopos} and \eqref{Eq:rhoneg1}, the rate of change of the dislocation density with respect to the plastic strain can be written as 
\begin{equation}{\label{Eq:Governing1}}
\frac{{\rm d}\rho}{{\rm d}\varepsilon_p}=\frac{\dot{\rho}_+-\dot{\rho}_-}{\dot{\varepsilon}_p}={2\pi\beta_1}\rho_{\rm source}\frac{\sigma-\sigma^*}{\rho \mu b^2}-\beta_2 \frac{M\rho}{2b\sqrt{\rho_{\rm source}}}.
\end{equation}

If the mean free path of the dislocations is limited by the film thickness, the rate of change of the dislocation density is given by
\begin{equation}{\label{Eq:Governing2}}
\frac{{\rm d}\rho}{{\rm d}\varepsilon_p}={2\pi\beta_1}\rho_{\rm source}\frac{\sigma-\sigma^*}{\rho \mu b^2}-\beta'_2 \frac{M}{bh}.
\end{equation}

Another length scale that may play a role in the governing equation is the spacing of obstacles in the film. In general, obstacles may be precipitates or immobile dislocations, and may result in the formation of pile-ups that increase the flow stress of the material. The hardening effect of obstacles is governed by their density and strengths \cite{Roos2001a,Chakravarthy2010a}. If the density of obstacles is large, the mean free paths $y_m$  may be limited by obstacle spacing and Eqs.~\eqref{Eq:Governing1} or \eqref{Eq:Governing2} may need to be modified.

Experiments and calculations \cite{Xiang2005,Xiang2006a,Nicola2006,Davoudi2012} indicate that the absence of passivation layers on the surfaces of a film lowers the flow stress and hardening rate.  If the surfaces of the film are not passivated, dislocations can escape from the film and a term that accounts for dislocations leaving the film needs to be subtracted from the right hand sides of Eqs.~\eqref{Eq:Governing1} or \eqref{Eq:Governing2}. This term is similar to the expression developed by Nix and Lee \cite{Nix2011a} for the rate of dislocations leaving micropillars, and is inversely proportional to the film thickness. 

As plastic flow proceeds, the dislocation density rises and the number of activated  sources increases. If the film contains a finite number of dislocation sources, $\rho_{\rm source}$ will approach the total density of sources and eventually the right hand side of Eq.~\eqref{Eq:Governing1}  will vanish: The evolution equation has an asymptote and the normal stress saturates. This behavior is observed in some experiments on bulk materials at high temperatures (see, e.g., Ref.~\cite{Kocks1975}) and in many simulations if the initial density of sources is high or the grains are relatively large (see, e.g., Ref.~\cite{Balint2008}). Higher temperatures, a high density of initial sources, and large grains ensure the validity of the Taylor equation with a constant coefficient. At high temperatures, dislocations become more dispersed; large grains delay the formation of pile-ups, and a high density of sources increases the interactions of dislocations on different slip planes compared to the interactions of dislocations on the same slip planes. When these conditions are met, the hardening rate decreases and the stress may reach a saturation stress provided that the governing equation~\eqref{Eq:Governing1}  remains true.

In this study, we consider a fixed number of dislocation sources. In three-dimensional problems, junctions form when two dislocations on different slip planes intersect. These junctions can restrict the motion of dislocations, provide pinning points, and act as new Frank-Read sources. It is then reasonable to assume that the density of activated  sources is proportional to the dislocation density \cite{Kocks1975}. If we insert $\rho_{\rm source}\propto\rho$  into Eq.~\eqref{Eq:Governing1} and use the Taylor equation~\eqref{Eq:Taylor}, we arrive at the deterministic  equation for the evolution of the total dislocation density proposed by H{\"a}hner \cite{Hahner1996a}.

The evolution equation developed in this paper contains two constants that need to be determined from experiments or simulations. Since dislocations do not leave the surface in the model considered here, the evolution equation can also be applied to bulk materials provided the correct density of sources is used. Thus it may be possible to determine these parameters from measurements performed on bulk materials with the same grain size. Alternately the parameters may be determined at the micro-scale using DDD simulations and then be used in a multiscale model for bulk materials.

\section{Conclusions}

Expressions have been developed for the density of activated dislocation sources, the rate of dislocation nucleation, and the rate of dislocation annihilation. These equations are summarized in Table~I. For low dislocation densities and very thin films, these expressions may change because of geometrical considerations. Comparison between the discrete dislocation simulations and the models reveals very good agreement. With the aid of these expressions, we have developed a dislocation evolution equation that contains just two parameters.

\begin{table}
\begin{center}
\begin{tabular}{| > {} p{3cm} |> {} l |}
\hline
{\footnotesize Density of activated dislocation sources} & {\footnotesize $\displaystyle \rho_{\rm source}= c \frac{\rho_{\rm source}^0}{2}\left[1+\frac{1}{N_{\rm slip\,sys}}\sum_{i=1}^{N_{\rm slip\,sys}} {\rm erf}\left(\frac{\sigma \sin \theta_i |\cos \theta_i|-\tau_{\rm avg}}{\sqrt{2}\tau_{\rm sd}}\right)\right]$} \\
\hline
{\footnotesize Rate of dislocation nucleation} & {\footnotesize $\displaystyle \frac{{\rm d}\rho_+}{{\rm d} t}=2\pi\beta_1\rho_{\rm source}\frac{\sigma-\sigma^*}{\rho \mu b^2}\frac{{\rm d} \varepsilon_p}{{\rm d} t}$} \\
\hline
{\footnotesize Rate of dislocation annihilation} & 
\begin{tabular}{l}
\\
{\footnotesize $\displaystyle \frac{{\rm d}\rho_-}{{\rm d}t}=\beta_2 \frac{M\rho}{2b\sqrt{\rho_{\rm source}}}\frac{{\rm d}\varepsilon_p}{{\rm d} t}$ \hspace{0.5cm} (for thicker films)}\\ 
\\
{\footnotesize $\displaystyle \frac{{\rm d}\rho_-}{{\rm d}t}= \frac{\beta'_2 M}{bh}\frac{{\rm d}\varepsilon_p}{{\rm d} t}$ \hspace{0.5cm} (for very thin films)}\\
\\
\end{tabular}
\\
\hline
{\footnotesize rate of change of the dislocation density with respect to  plastic strain} & 
\begin{tabular}{l}
\\
{\footnotesize  $\displaystyle  \frac{{\rm d}\rho}{{\rm d}\varepsilon_p}={2\pi\beta_1}\rho_{\rm source}\frac{\sigma-\sigma^*}{\rho \mu b^2}-\beta_2 \frac{M\rho}{2b\sqrt{\rho_{\rm source}}} $  \hspace{0.5cm} (for thicker films)}\\
\\
{\footnotesize    $\displaystyle \frac{{\rm d}\rho}{{\rm d}\varepsilon_p}={2\pi\beta_1}\rho_{\rm source}\frac{\sigma-\sigma^*}{\rho \mu b^2}-\beta'_2 \frac{M}{bh} $ \hspace{0.5cm} (for very think films)}
\end{tabular}
\\
\hline
\end{tabular}
\end{center}
\end{table}

\section*{Acknowledgements}

The authors gratefully acknowledge support from NSF (Grant DMR-0820484). The authors wish to thank Professor Lucia Nicola of Delft University for the help with the DDD simulation code and for insightful discussions. 
 
\begin{small}
{\sffamily

}
\end{small}

\newpage

\appendix
\numberwithin{equation}{section}
\numberwithin{figure}{section}

\section{Taylor Equation}{\label{Section:TaylorEquation}}

The Taylor equation was one of the first expressions relating the flow stress of a material to its dislocation density. The expression was first developed by G.I. Taylor \cite{Taylor1934} in an attempt to describe work hardening. The equation arises naturally if one assumes the flow stress is the external stress required to drive two dislocations on parallel slip planes past one another \cite{Messerschmidt2010}. Given that the maximum shear stress associated with a dislocation is of order $\mu b/r$, where $\mu$ is the shear modulus and $r$ the distance to the dislocation, and that the average spacing between randomly distributed dislocations is of order $1/\sqrt{\rho}$, the flow stress $\tau$ of a material can be written as
\begin{equation}\label{Eq:one}
\tau=\tau_0+\alpha \mu b \sqrt{\rho}.
\end{equation}
In this expression, $\alpha$ is a dimensionless parameter ranging from 0.05 to 2.6 for different materials \cite{Lavrentev1980}, and $\tau_0$ is the flow stress of the material in the absence of dislocation interactions. In other words, $\tau_0$ is the shear resistance to dislocation motion when $\rho\approx 0$ \cite{Lavrentev1980}.  Other work-hardening models lead to a similar linear relation between the flow stress and the square root of the dislocation density, but with different proportionality constants \cite{Nabarro1964}. It is convenient to rewrite Eq.~(\ref{Eq:one}) as in \cite{Viguier2003}
\begin{equation}{\label{Eq:Taylortau2}}
\tau-\tau^*=\alpha \mu b \left(\sqrt{\rho}-\sqrt{\rho^*}\right),
\end{equation}
where $\tau^*$ and $\rho^*$ are the flow stress and dislocation density at the point where the material first becomes plastic. Many experiments have shown that the Taylor equation holds true for f.c.c., b.c.c., and h.c.p metals, as well as for ionic and covalent materials \cite{Viguier2003}, both in single crystals and in polycrystals, as long as the flow stress is solely controlled by interactions between dislocations \cite{Kocks2003}. The shear flow stress $\tau$ of a single crystal to the uniaxial flow stress $\sigma$  of a polycrystal are related through $\sigma=M\tau$, where in our model $M=3.1$ is equivalent to the Taylor factor. In the absence of a crystallographic texture, the Taylor factor takes a value of 3.1 for f.c.c. and b.c.c. metals in tension or compression \cite{Hull2011,Kocks1970}. Therefore, the Taylor equation can be reformulated for a uniaxial loading of a polycrystal as:

\begin{equation}{\label{Eq:Taylor}}
\sigma=\sigma^*+M\alpha b \mu \left(\sqrt{\rho}-\sqrt{\rho^*}\right).
\end{equation}

\begin{figure}[h!]
\centering
\begin{tabular}{p{0.05cm}p{8cm}}
\vspace{10pt} {(a)} & 
\vspace{0pt} \includegraphics[width=8cm]{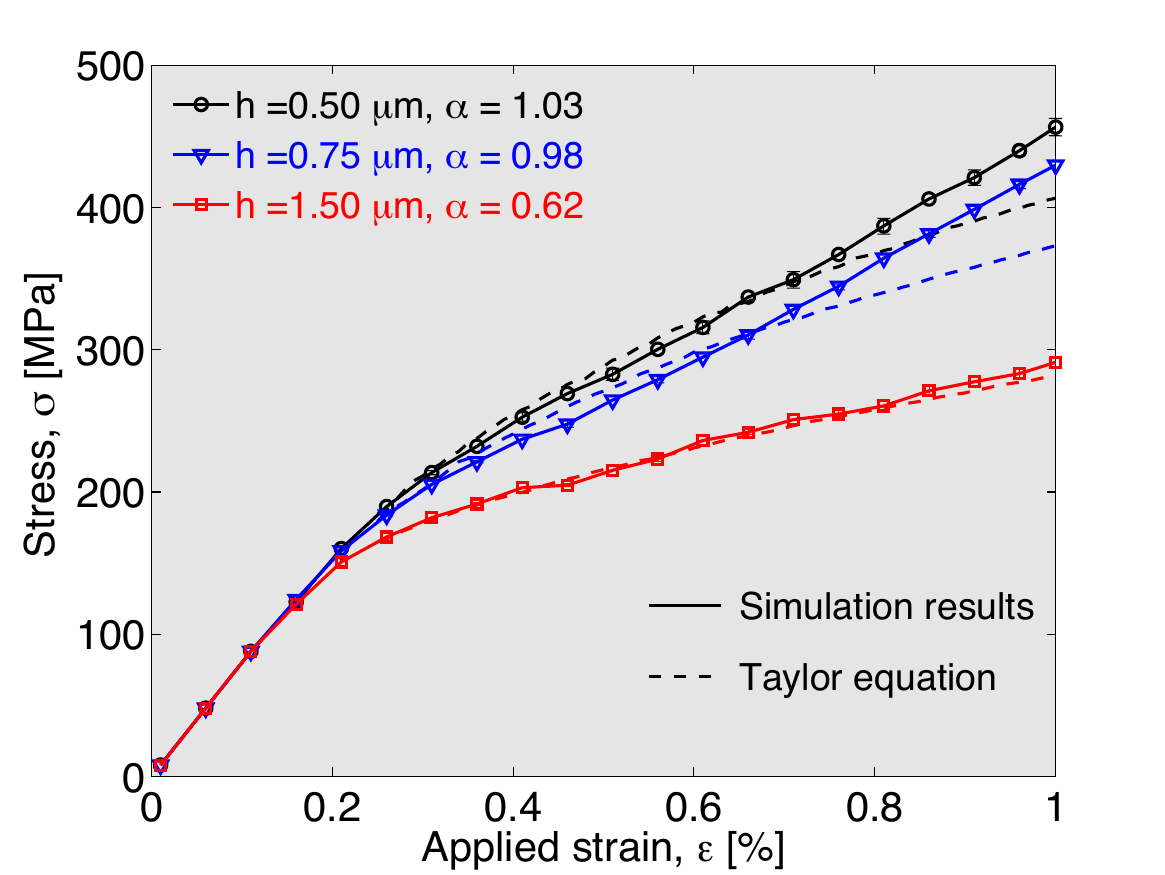} \\
\vspace{10pt} {(b)} & 
\vspace{0pt}  \includegraphics[width=8cm]{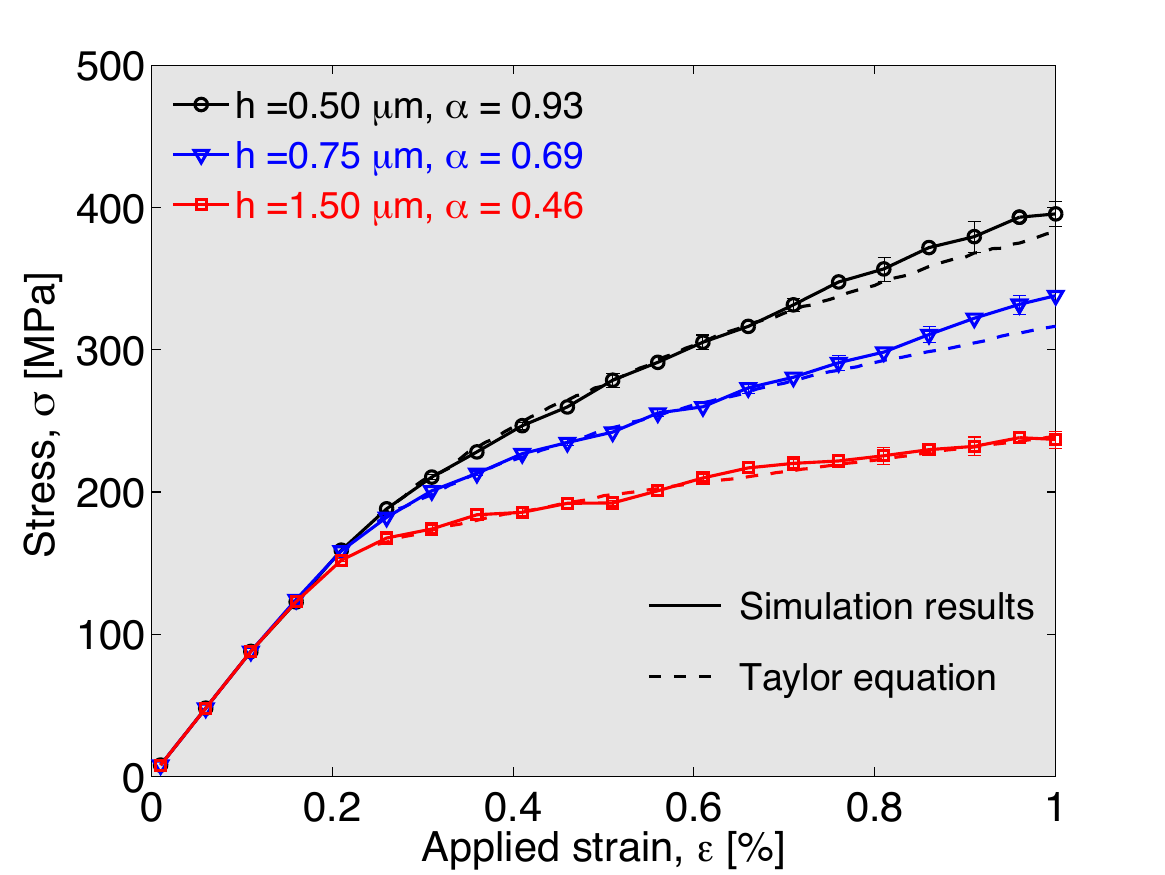} 
\end{tabular}
\caption{\label{Fig:TaylorEqFigure} The stress-strain curves are plotted for three different film thicknesses $h$ when dislocations (a) can only glide and (b) can glide and climb.  Error bars show the standard error. Comparison between simulation and model shows that the Taylor equation is satisfied up for strains smaller than 0.7 \%. } 
\end{figure}

{\color{black}
Figure~\ref{Fig:TaylorEqFigure} shows several stress-strain curves obtained for films with different thicknesses using DDD simulations. The solid lines represent the simulation results; the dashed lines represent the stress-strain curves derived from the Taylor model, Eq.~(\ref{Eq:Taylor}), using the results of the simulations in the following manner: at each strain, the stress and the dislocation density are known from the simulation results. The proportionality constant is determined by linear regression of  $\sigma-\sigma^*$ on $M\mu b\left(\sqrt{\rho}-\sqrt{\rho^*}\right)$. Then $\alpha M\mu b\left(\sqrt{\rho}-\sqrt{\rho^*}\right)$ versus the strain for different film thicknesses are depicted by dashed lines. The figure clearly illustrates that the Taylor equation provides a good fit to the simulation data for small strains ($\varepsilon<$0.7\%), whether or not dislocation climb is enabled. 
}

When dislocation climb is enabled, dislocations are more dispersed, and the average spacing between dislocations is larger than when dislocations can only glide. For this reason, the values of $\alpha$ are smaller in Fig.~\ref{Fig:TaylorEqFigure}(b) than in Fig.~\ref{Fig:TaylorEqFigure}(a). A large number of experimental observations indicate that $\alpha$ decreases with increasing temperature \cite{Kocks2003}. This decrease is attributed to the activation of recovery mechanisms such as cross slip and dislocation climb and to the dispersion of dislocations, in line with what is observed here.

At larger strains, the agreement between the Taylor equation and the stress-strain curves in Fig.~\ref{Fig:TaylorEqFigure}(a) is not as satisfying: the stress-train curves derived from the simulations tend to show linear hardening, while the Taylor equation predicts parabolic hardening. This discrepancy can be attributed to the formation of dislocation pile-ups in the simulation. As plastic deformation proceeds, the number of dislocations in pile-ups increases and the number of dislocations is a linear function of the external stress \cite{Leibfried1951,Hirth1982}.

When the dislocation structure is converted from a random distribution to an organized microstructure, Eq.~(\ref{Eq:Taylor}) remains valid if $\alpha$ is allowed to vary with $\varepsilon_p$. For example, experiments on bulk materials have shown that $\alpha$ decreases with increasing deformation when dislocation cells form inside the grains \cite{Kocks2003}.  The results in Fig.~\ref{Fig:TaylorEqFigure} suggest that pile-up formation in thin films may cause $\alpha$ to increase with $\varepsilon_p$. If $\alpha$ varies with $\varepsilon_p$, the work-hardening rate $\theta_p$ of a polycrystalline metal can be written as

\begin{equation}
\theta_p\equiv\frac{{\rm d}\sigma}{{\rm d}\varepsilon_p}=\frac{\alpha \mu M b}{2\sqrt{\rho}}\frac{{\rm d}\rho}{{\rm d}\varepsilon_p}+\left(\sigma-\sigma^*\right)\frac{{\rm d}\ln \alpha}{{\rm d}\varepsilon_p}.
\end{equation}
The change in $\alpha$ is usually negligible for small strains. Thus finding the hardening rate requires an expression for $\textrm{d}\rho/\rm{d}\varepsilon_p$.


\begin{thebibliography}{10}

\bibitem{Cottrell2002}
A. H. Cottrell.
\newblock {Commentary. A brief view of work hardening}.
\newblock In {\em Dislocations in solids 11}, pages vii--xvii. Elsevier B.V. (2002).

\bibitem{Kubin2009}
L. P. Kubin, B. Devincre, and T. Hoc.
\newblock {The deformation stage II of face-centered cubic crystals: Fifty
  years of investigations}.
\newblock {\em International Journal of Materials Research},
  100(10):1411--1419 (2009).

\bibitem{Kocks2003}
U. F. Kocks and H~Mecking.
\newblock {Physics and phenomenology of strain hardening: the FCC case}.
\newblock {\em Progress in Materials Science}, 48(3):171--273 (2003).

\bibitem{Kubin1992}
L. P. Kubin and G.~Canova.
\newblock {The modelling of dislocation patterns}.
\newblock {\em Scripta Metallurgica et Materialia}, 27(8):957--962 (1992).

\bibitem{Sauzay2011}
M. Sauzay and L. P. Kubin.
\newblock {Scaling laws for dislocation microstructures in monotonic and cyclic
  deformation of fcc metals}.
\newblock {\em Progress in Materials Science}, 56(6):725--784 (2011).

\bibitem{Hillert2006}
M. Hillert and J. {\AA}gren.
\newblock {Extremum principles for irreversible processes}.
\newblock {\em Acta Materialia}, 54(8):2063--2066 (2006).

\bibitem{Kocks1975}
U. F. Kocks, A. S. Argon, and M. F. Ashby.
\newblock {\em {Thermodynamics and kinetics of slip}}.
\newblock Pergamon Press Ltd. (1975).

\bibitem{Mecking1981}
H. Mecking and U. F. Kocks.
\newblock {Kinetics of flow and strain-hardening}.
\newblock {\em Acta Metallurgica}, 29:1865--1875 (1981).

\bibitem{Estrin1984}
Y. Estrin and H. Mecking.
\newblock {A unified phenomenological description of work hardening and creep
  based on one-parameter models}.
\newblock {\em Acta Metallurgica}, 32(I):57--70 (1984).

\bibitem{Walgraef1985b}
D. Walgraef and E. C. Aifantis.
\newblock {On the formation and stability of dislocation patterns -I:
  one-dimensional considerations}.
\newblock {\em International journal of engineering science}, 23:1351--1358 (1985).

\bibitem{Walgraef1985}
D. Walgraef and E. C. Aifantis.
\newblock {On the formation and stability of dislocation patterns -II:
  two-dimensional considerations}.
\newblock {\em International journal of engineering science},
  23(12):1359--1364 (1985).

\bibitem{Walgraef1985a}
D. Walgraef and E. C. Aifantis.
\newblock {On the formation and stability of dislocation patterns -III:
  three-dimensional considerations}.
\newblock {\em International journal of engineering science},
  23(12):1365--1372 (1985).

\bibitem{Aifantis1986}
E. C. Aifantis.
\newblock {On the Dynamical Origin of Dislocation Patterns}.
\newblock {\em Materials Science and Engineering}, 81:563--574 (1986).

\bibitem{Follansbee1988}
P. S. Follansbee and U. F. Kocks.
\newblock {A constitutive description of the deformation of copper based on the
  use of the mechanical threshold stress as an internal state variable}.
\newblock {\em Acta Metallurgica}, 36(1):81--93 (1988).

\bibitem{Hahner1996a}
P. H{\"{a}}hner.
\newblock {A theory of dislocation cell formation on stochastic dislocation
  dynamics based}.
\newblock {\em Acta materialia}, 44(6):2345--2352 (1996).

\bibitem{Hahner1996}
P. H{\"{a}}hner.
\newblock {On the foundations of stochastic dislocation dynamics}.
\newblock {\em Applied Physics A Materials Science and Processing},
  62(5):473--481 (1996).

\bibitem{Nix2011a}
W. D. Nix and S. W. Lee.
\newblock {Micro-pillar plasticity controlled by dislocation nucleation at
  surfaces}.
\newblock {\em Philosophical Magazine}, 91:1084--1096 (2011).

\bibitem{Devincre2008}
B. Devincre, T. Hoc, and L. P. Kubin.
\newblock {Dislocation mean free paths and strain hardening of crystals.}
\newblock {\em Science}, 320(5884):1745--1748 (2008).

\bibitem{Ryu2013a}
I. Ryu, W. D. Nix, and W. Cai.
\newblock {Plasticity of bcc micropillars controlled by competition between
  dislocation multiplication and depletion}.
\newblock {\em Acta Materialia}, 61(9):3233--3241 (2013).

\bibitem{Sills2016}
R. B. Sills, A. Aghaei, and W. Cai.
\newblock {Advanced time integration algorithms for dislocation dynamics
  simulations of work hardening}.
\newblock {\em Modelling and Simulation in Materials}, 24(4) 045019 (2016).

\bibitem{Benzerga2004a}
A. A. Benzerga, Y. Br{\'{e}}chet, A. Needleman, and E. {Van der Giessen}.
\newblock {Incorporating three-dimensional mechanisms into two-dimensional
  dislocation dynamics}.
\newblock {\em Modelling and Simulation in Materials Science and Engineering},
  12(1):159--196 (2004).

\bibitem{Gomez-Garcia2006}
D. G{\'{o}}mez-Garc{\'{i}}a, B. Devincre, and L. P. Kubin.
\newblock {Dislocation Patterns and the Similitude Principle: 2.5D Mesoscale
  Simulations}.
\newblock {\em Physical Review Letters}, 96(12):8--11 (2006).

\bibitem{Keralavarma2016}
S. M. Keralavarma and W. A. Curtin.
\newblock {Journal of the Mechanics and Physics of Solids Strain hardening in
  2D discrete dislocation dynamics simulations : A new '2 .5D' algorithm}.
\newblock {\em Journal of the Mechanics and Physics of Solids}, 95:132--146 (2016).

\bibitem{Ispanovity2010}
P. D. Isp{\'{a}}novity, I. Groma, G.
  Gy{\"{o}}rgyi, F. F. Csikor, and D. Weygand.
\newblock {Submicron Plasticity: Yield Stress, Dislocation Avalanches, and
  Velocity Distribution}.
\newblock {\em Physical Review Letters}, 105(8):085503 (2010).

\bibitem{Ispanovity2013}
P. D. Isp{\'{a}}novity, {\'{A}}. Hegyi, I. Groma,
  G. Gy{\"{o}}rgyi, K. Ratter, and D. Weygand.
\newblock {Average yielding and weakest link statistics in micron-scale
  plasticity}.
\newblock {\em Acta Materialia}, 61(16):6234--6245 (2013).

\bibitem{Agnihotri2015}
P. K. Agnihotri and E. {Van Der Giessen}.
\newblock {On the rate sensitivity in discrete dislocation plasticity}.
\newblock {\em Mechanics of Materials}, 90:37--46 (2015).

\bibitem{Keralavarma2015}
S. M. Keralavarma and A. A. Benzerga.
\newblock {High-temperature discrete dislocation plasticity}.
\newblock {\em Journal of the Mechanics and Physics of Solids}, 82: 1--22 (2015).

\bibitem{Shishvan2016a}
S. S. Shishvan, T. M. Pollock, R. M. McMeeking, and V. S.
  Deshpande.
\newblock {Interfacial diffusion in high-temperature deformation of composites:
  a discrete dislocation plasticity investigation}.
\newblock {\em Journal of the Mechanics and Physics of Solids}, 98:330--349 (2017).

\bibitem{Gurrutxaga-Lerma2017}
B. Gurrutxaga-lerma, D. S. Balint, D. Dini, and A. P. Sutton.
\newblock {A Dynamic Discrete Dislocation Plasticity study of elastodynamic
  shielding of stationary cracks}.
\newblock {\em Journal of the Mechanics and Physics of Solids},
  98:1--11 (2017).

\bibitem{Quek2016}
S. S. Quek, Z. H. Chooi, Z. Wu, Y. W. Zhang, and D. J.
  Srolovitz.
\newblock {The inverse hall-petch relation in nanocrystalline metals: A
  discrete dislocation dynamics analysis}.
\newblock {\em Journal of the Mechanics and Physics of Solids}, 88:252--266 (2016).

\bibitem{El-awady2015a}
J. A. El-awady.
\newblock{dislocation-mediated plasticity.}
\newblock {\em Nature Communications}, 6:5926 (2015).

\bibitem{Madec2017}
R. Madec and L. P. Kubin.
\newblock {Dislocation strengthening in FCC metals and in BCC metals at high
  temperatures}.
\newblock {\em Acta Materialia}, 126:166--173 (2017).

\bibitem{VanderGiessen1995}
E. {Van der Giessen} and A. Needleman.
\newblock {Discrete dislocation plasticity : a simple planar model}.
\newblock {\em Modelling and Simulation in Materials Science and Engineering},
  3:689--735 (1995).

\bibitem{Kubin1992a}
L. P. Kubin, G. Canova, M. Condat, B. Devincre, V. Pontikis, and
  Y. Br{\'{e}}chet.
\newblock {Dislocation Microstructures and Plastic Flow: A 3D Simulation}.
\newblock {\em Solid State Phenomena}, 23:455--472 (1992).

\bibitem{Davoudi2017}
K. M. Davoudi.
\newblock {Temperature dependence of the yield strength of aluminum thin films:
  Multiscale modeling approach}.
\newblock {\em Scripta Materialia}, 131:63--66 (2017).

\bibitem{Bulatov2006a}
V. V. Bulatov and W. Cai.
\newblock {\em computer simulations of dislocations}.
\newblock Oxford University Press (2006).

\bibitem{Messerschmidt2010}
U. Messerschmidt.
\newblock {\em {Dislocation dynamics during plastic deformation}}.
\newblock Springer-Verlag, Berlin (2010).

\bibitem{Weertman1955}
J. Weertman.
\newblock {Theory of Steady-State Creep Based on Dislocation Climb}.
\newblock {\em Journal of Applied Physics}, 26(10):1213 (1955).

\bibitem{Schoeck1957}
G. Schoeck.
\newblock {Theory of Creep}.
\newblock In {\em Creep and Recovery}, pages 199--226, Cleveland, Ohio (1957).
  American Society for Metals.

\bibitem{Schoeck1980}
G. Schoeck.
\newblock {Thermodynamics and thermal activation of dislocations}.
\newblock In F.~R.~N. Nabarro, editor, {\em Dislocations in solids, vol 3},
  pages 63--163. North-Holland (1980).

\bibitem{Davoudi2012}
K. M. Davoudi, L. Nicola, and J. J. Vlassak.
\newblock {Dislocation climb in two-dimensional discrete dislocation dynamics}.
\newblock {\em Journal of Applied Physics}, 111(10):103522 (2012).

\bibitem{Davoudi2014}
K. M. Davoudi, L. Nicola, and J. J. Vlassak.
\newblock {Bauschinger effect in thin metal films: Discrete dislocation
  dynamics study}.
\newblock {\em Journal of Applied Physics}, 115(1):013507 (2014).

\bibitem{Ayas2012a}
C. Ayas, V. S. Deshpande, and M. G. D. Geers.
\newblock {Tensile response of passivated films with climb-assisted dislocation
  glide}.
\newblock {\em Journal of the Mechanics and Physics of Solids}, 60:1626--1643 (2012).

\bibitem{Ayas2015}
C. Ayas, L. C. P. Dautzenberg, M. G. D. Geers, and V. S. Deshpande.
\newblock {Climb-Enabled Discrete Dislocation Plasticity Analysis of the
  Deformation of a Particle Reinforced Composite}.
\newblock {\em Journal of Applied Mechanics}, 82(7):071007 (2015).

\bibitem{Nicola2006}
L. Nicola, Y. Xiang, J. J. Vlassak, E. {Van der Giessen}, and A.
  Needleman.
\newblock {Plastic deformation of freestanding thin films: Experiments and
  modeling}.
\newblock {\em Journal of the Mechanics and Physics of Solids},
  54(10):2089--2110 (2006).

\bibitem{Rice1987}
J. R. Rice.
\newblock {Tensile crack tip fields in elastic-ideally plastic crystals}.
\newblock {\em Mechanics of Materials}, 6(4):317--335 (1987).

\bibitem{Taylor1934}
G. I. Taylor.
\newblock {The Mechanism of Plastic Deformation of Crystals. Part
  I.-Theoretical}.
\newblock {\em Proceedings of the Royal Society of London. Series A, Containing
  Papers of a Mathematical and Physical Character}, 145(855):362--387 (1934).

\bibitem{Kubin2013}
L. P. Kubin.
\newblock {\em dislocations, mesoscale simulations and plastic flow}.
\newblock Oxford University Press (2013).

\bibitem{Roos2001a}
A. Roos, J. Th. M. {De Hosson}, and E. {Van der Giessen}.
\newblock {A two-dimensional computational methodology for high-speed
  dislocations in high strain-rate deformation}.
\newblock {\em Computational Materials Science}, 20(1):1--18 (2001).

\bibitem{Chakravarthy2010a}
S. S. Chakravarthy and W. A. Curtin.
\newblock {Effect of source and obstacle strengths on yield stress: A discrete
  dislocation study}.
\newblock {\em Journal of the Mechanics and Physics of Solids}, 58(5):625--635 (2010).

\bibitem{Xiang2005}
Y. Xiang and J. J. Vlassak.
\newblock {Bauschinger effect in thin metal films}.
\newblock {\em Scripta Materialia}, 53(2):177--182 (2005).

\bibitem{Xiang2006a}
Y. Xiang and J. J. Vlassak.
\newblock {Bauschinger and size effects in thin-film plasticity}.
\newblock {\em Acta Materialia}, 54(20):5449--5460 (2006).

\bibitem{Balint2008}
D. S. Balint, V. S. Deshpande, A. Needleman, and E. {Van der
  Giessen}.
\newblock {Discrete dislocation plasticity analysis of the grain size
  dependence of the flow strength of polycrystals}.
\newblock {\em International Journal of Plasticity}, 24(12):2149--2172 (2008).

\bibitem{Lavrentev1980}
F. F Lavrentev.
\newblock {The type of dislocation interaction as the factor determining work
  hardening}.
\newblock {\em Materials Science and Engineering}, 46:191--208 (1980).

\bibitem{Nabarro1964}
F. R. N. Nabarro, Z. S. Basinski, and D. B. Holt.
\newblock {The plasticity of pure single crystals}.
\newblock {\em Advances in Physics}, 13(50):193--323 (1964).

\bibitem{Viguier2003}
B. Viguier.
\newblock {Dislocation densities and strain hardening rate in some
  intermetallic compounds}.
\newblock {\em Materials Science and Engineering: A}, 349:132--135 (2003).

\bibitem{Hull2011}
Derek Hull and David~J. Bacon.
\newblock {\em {Introduction to dislocations}}.
\newblock Elsevier, 5 edition, 2011.

\bibitem{Kocks1970}
U. F. Kocks.
\newblock {The relation between polycrystal deformation and single-crystal
  deformation}.
\newblock {\em Metallurgical and Materials Transactions}, 1:1121--1143 (1970).

\bibitem{Leibfried1951}
G. Leibfried.
\newblock {Verteilung von Versetzungen im statischen Gleichgewicht}.
\newblock {\em Zeitschrift f{\"u}r Physik}, 130:214--226 (1951).

\bibitem{Hirth1982}
J. P. Hirth and J. Lothe.
\newblock {\em {Theory of dislocations}}.
\newblock John Wiley {\&} Sons, 2nd edition (1982).

\end{thebibliography}
\end{document}